\documentclass[a4paper,runningheads]{llncs} 
\usepackage[T1]{fontenc}
\usepackage{lmodern}
\usepackage{amsmath,amsfonts}
\usepackage{graphicx}
\usepackage{url}
\usepackage{xcolor}
\usepackage{booktabs}
\usepackage{float}
\usepackage{caption}
\usepackage{subcaption}
\usepackage{hyperref} 
\usepackage{tikz}
\usetikzlibrary{
    positioning,
    shapes,
    arrows.meta,
    shapes.geometric,
    calc,
    decorations.pathmorphing,
    shadows,
    patterns,
    shapes.misc,
    matrix,
    backgrounds,
    fit
}
\definecolor{perceptionblue}{HTML}{2980B9}
\definecolor{edgegreen}{HTML}{27AE60}
\definecolor{omniverseorange}{HTML}{F39C12}
\definecolor{displaypurple}{HTML}{8E44AD}
\definecolor{flred}{HTML}{C0392B}
\definecolor{datagray}{HTML}{7F8C8D}
\tikzset{
  panel/.style={draw, thick, minimum width=5cm, minimum height=3.5cm, fill=gray!10, rounded corners},
  title/.style={font=\bfseries\sffamily, anchor=north west, xshift=0.1cm, yshift=-0.1cm},
  label_text/.style={font=\sffamily\scriptsize, align=center},
  arrow/.style={->, thick}
}
\newcommand{\icon}[1]{#1} 
\begin{document}

\title{Holo-Artisan: A Personalized Multi-User Holographic Experience for Virtual Museums on the Edge Intelligence}

\titlerunning{Holo-Artisan: Personalized Holographic Experiences on the Edge Intelligence}

\author{Nan-Hong Kuo\inst{1} \and
Hojjat Baghban\inst{2}}
\authorrunning{N.-H. Kuo et al.}
\institute{National Taiwan University, Taipei, Taiwan \and
Chang Gung University, Taoyuan, Taiwan\\
\email{\{d91222008@ntu.edu.tw, hojjat.baghban@cgu.edu.tw\}}}

\maketitle

\begin{abstract}
\fontsize{10}{12}\selectfont
\noindent
We present Holo-Artisan, a novel system architecture enabling immersive multi-user experiences in virtual museums through true holographic displays and personalized edge intelligence. In our design, local edge computing nodes process real-time user data – including pose, facial expression, and voice, for multiple visitors concurrently. Generative AI models then drive digital artworks (e.g., a volumetric Mona Lisa) to respond uniquely to each viewer. For instance, the Mona Lisa can return a smile to one visitor while engaging in a spoken Q\&A with another, all in real time. A cloud-assisted collaboration platform composes these interactions in a shared scene using a universal scene description, and employs ray tracing to render high-fidelity, personalized views with a direct pipeline to glasses-free holographic displays. To preserve user privacy and continuously improve personalization, we integrate federated learning (FL) – edge devices locally fine-tune AI models and share only model updates for aggregation. This edge-centric approach minimizes latency and bandwidth usage, ensuring a synchronized shared experience with individual customization. Through Holo-Artisan, static museum exhibits are transformed into dynamic, \emph{living} artworks that engage each visitor in a personal dialogue, heralding a new paradigm of cultural heritage interaction.
\end{abstract}

\section{Introduction}

Immersive technologies are revolutionizing how we experience cultural heritage. However, existing virtual museum platforms often present static, pre-rendered content, failing to provide the dynamic, multi-user, and personalized interactions needed for true engagement. A shared digital artifact, such as a volumetric representation of the Mona Lisa, cannot currently react in real-time to the individual behaviors of multiple simultaneous viewers---it cannot smile back at one visitor while answering another's question. This limitation creates a passive, one-size-fits-all experience.

This paper introduces Holo-Artisan, a novel system architecture designed to bridge this gap. We integrate the collaborative power of 3D collaboration platforms, the responsiveness of Edge AI, the privacy-preserving capabilities of federated learning (FL), and the realism of generative AI to create "living" artworks. Our system enables multiple users in a shared virtual space to interact with a digital exhibit, receiving unique, personalized responses while maintaining a coherent shared reality.

\begin{figure*}[htbp]
    \centering
     \begin{subfigure}[b]{0.38\textwidth}
        \centering
        \resizebox{\textwidth}{!}{%
        \begin{tikzpicture}[
            font=\sffamily\small,
            user/.style={circle, draw=perceptionblue, fill=perceptionblue!20, thick, minimum size=1.5cm, align=center},
            artwork/.style={rectangle, rounded corners, draw=omniverseorange, fill=omniverseorange!10, thick, minimum width=4.5cm, minimum height=5.5cm, align=center, drop shadow},
            response/.style={ellipse, draw=edgegreen, fill=edgegreen!10, thick, minimum width=2.5cm, align=center},
            arrow/.style={-Latex, thick},
            dashed_arrow/.style={-Latex, thick, dashed, datagray}
        ]
            \node[artwork] (base_mona) at (0,0) {
                \includegraphics[width=2.5cm]{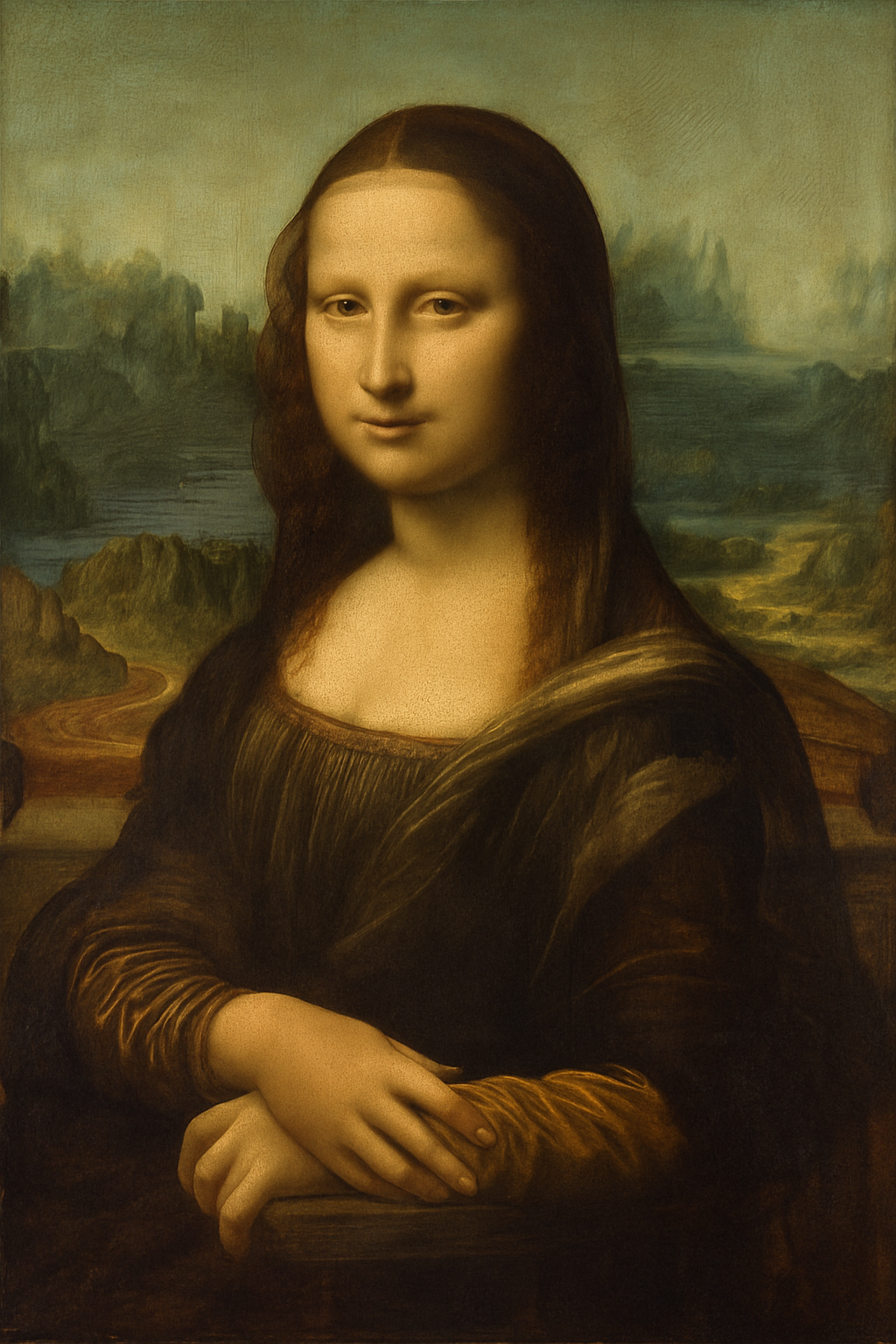} \\ 
                \textbf{Shared Base Model} \\
                \small Mona Lisa (Neutral State) \\
                \tiny (Scene Description File)
            };
        
            \node[user] (user1) at (-5, 2.5) {\textbf{User 1}};
            \node[response, below=0.7cm of user1] (view1) {Sees a\\Subtle Smile};
            \draw[arrow, perceptionblue] (user1) to[bend left=15] node[midway, above, sloped] {Input: Smile Detected} (base_mona.north west);
            \draw[dashed_arrow, edgegreen] (base_mona.west) to[bend left=15] node[midway, below, sloped] {Personalized View 1} (view1);
        
            \node[user] (user2) at (5, 2.5) {\textbf{User 2}};
            \node[response, below=0.7cm of user2] (view2) {Sees \& Hears\\a Spoken Reply};
            \draw[arrow, perceptionblue] (user2) to[bend right=15] node[midway, above, sloped] {Input: Voice Question} (base_mona.north east);
            \draw[dashed_arrow, edgegreen] (base_mona.east) to[bend right=15] node[midway, below, sloped] {Personalized View 2} (view2);
        
            \node[user] (user3) at (0, -6) {\textbf{User 3}};
            \node[response, above=0.7cm of user3] (view3) {Sees the\\Neutral Portrait};
            
            \draw[arrow, perceptionblue] (user3) to node[midway, left=3em] {Input: Passive View} (base_mona.south);
            \draw[dashed_arrow, edgegreen] (base_mona.south) to[bend right=180] node[midway, right=2em] {Consistent Base View} (view3);
        
        \end{tikzpicture}
        }
        \caption{Conceptual illustration of multi-user interaction.}
        \label{fig:mona_lisa_interaction_concept}
        \end{subfigure}
    \hfill
    \begin{subfigure}[b]{0.6\textwidth}
        \centering
        \resizebox{\textwidth}{!}{%
        \begin{tikzpicture}[
            font=\normalsize,
            node distance=1cm,
            lifeline/.style={gray!80, thick},
            activity/.style={rectangle, rounded corners=2pt, draw=black, fill=white, thick, minimum height=1.3cm, minimum width=0.9cm},
            message/.style={-Latex, thick},
            note/.style={rectangle, draw=black!50, fill=yellow!10, text width=4.0cm, align=left, font=\normalsize},
            flred/.style={red!70!black}
        ]
            \node (user_label)       at (0,0)  {User};
            \draw[lifeline] (0,-0.5) -- (0,-11);
    
            \node (perc_label)       at (3,0)  {Perception Layer};
            \draw[lifeline] (3,-0.5) -- (3,-11);
    
            \node (edge_label)       at (6,0)  {Edge Intelligence};
            \draw[lifeline] (6,-0.5) -- (6,-11);
    
            \node (fl_label)         at (9,0)  {FL Aggregator};
            \draw[lifeline] (9,-0.5) -- (9,-11);
    
            \node (rend_label)       at (12,0) {Rendering Layer};
            \draw[lifeline] (12,-0.5) -- (12,-10.5);
    
            \node (disp_label)       at (15,0) {Display};
            \draw[lifeline] (15,-0.5) -- (15,-11);
    
            \node[activity] (user_act) at (0,-1.5)  {};
            \draw[message] (user_act.east) -- (3,-1.5) node[midway, above] {Gesture/Voice};
    
            \node[activity] (perc_act) at (3,-2.6) {};
            \draw[message] (3,-2.1) -- (perc_act.north);
            \draw[message] (perc_act.east) -- (6,-2.6) node[midway, above] {User State $U_i(t)$};
    
            \node[activity] (edge_act) at (6,-4)  {};
            \draw[message] (6,-3.3) -- (edge_act.north);
            \node[note, right=0.4cm of edge_act, yshift=0.6cm] {1. Perform AI inference\\2. Generate response cmd\\3. Log data for training};
    
            \draw[message] (edge_act.east) -- (12,-6.6) node[midway, above] {Response Cmd $R_i$};
    
            \node[activity] (rend_act) at (12,-7.2) {};
            \draw[message] (12,-6.6) -- (rend_act.north);
            \draw[message] (rend_act.east) -- (15,-7.2) node[midway, above] {Personalized Frame};
    
            \node[activity] (disp_act) at (15,-8.2) {};
            \draw[message, dashed] (disp_act.west) -- (0,-8.2) node[midway, above] {Visual/Audio Feedback};
    
            \draw[message, flred, dashed] (edge_act.south) to[bend right=12] (9,-9.0) node[midway, above, sloped, yshift=0.5cm] {\small Upload Model Update $\Delta w_i$};
    
            \node[activity, fill=red!10, minimum width=1.4cm, minimum height=1.8cm, font=\small\bfseries] (fl_agg) at (9,-9.8) {FL\newline Agg};
            \node[note, left=0.5cm of fl_agg] {Periodically aggregates updates from all active edge nodes.};
    
            \draw[message, flred, dashed] (fl_agg.south) to[bend right=12] (6,-11) node[midway, below, sloped, yshift=-0.3cm] {\small Download Global Model $w_{t+1}$};
        \end{tikzpicture}}
        \caption{Sequence diagram of the interaction and learning loop.}
        \label{fig:SequenceDiagram}
    \end{subfigure}
    \caption{Conceptual (a) and sequential (b) diagrams of the Holo-Artisan interaction flow. The system architecture enables personalized responses to different users from a shared model, following a low-latency pathway for real-time feedback and an asynchronous loop for privacy-preserving model updates.}
    \label{fig:interaction_diagrams}
\end{figure*}
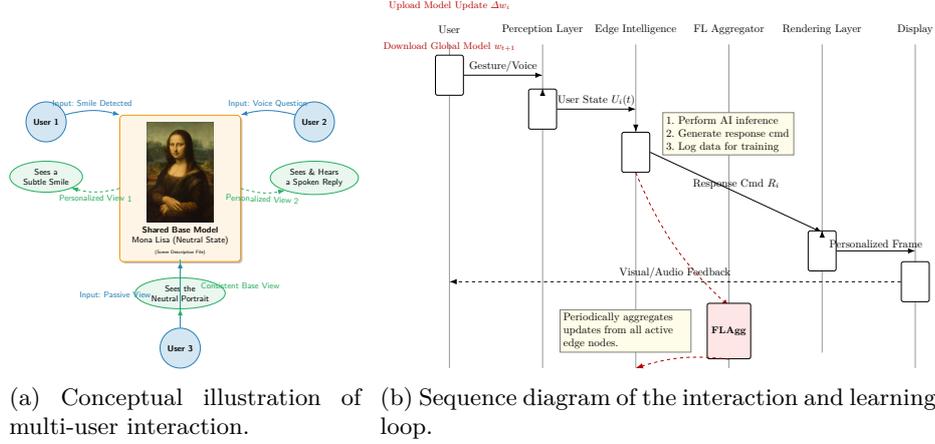

Imagine two visitors, Alice and Bob, in our virtual museum, standing before the holographic Mona Lisa. When Alice smiles, the artwork subtly smiles back only to her. Simultaneously, when Bob asks a question aloud, he sees and hears the portrait provide a spoken answer, with its lips synchronized to the words he hears. A third passive observer sees only the original, neutral portrait. This seamless, personalized interaction, illustrated conceptually in Figure~\ref{fig:interaction_diagrams}, is made possible by our distributed architecture where local AI models on edge devices drive personalization, while a central platform composes these unique responses into a shared base scene. User data remains on-device, with only anonymous model updates being shared via a federated learning framework to continuously improve the personalization engine while preserving privacy.

The main contributions of this work are:
\begin{enumerate}
    \item A novel system architecture that synergizes edge computing, federated learning, generative AI, and a 3D collaboration platform for personalized, multi-user interactions in virtual museums.
    \item The design of a personalized response engine that dynamically renders visual and auditory reactions based on real-time user perception.
    \item The establishment of a theoretical framework to analyze the system's latency, bandwidth, and consistency, highlighting expected performance gains over traditional cloud-based approaches.
    \item A proof-of-concept design, centered on the iconic Mona Lisa, showcasing the potential to transform cultural engagement from passive observation to active, personal dialogue.
\end{enumerate}

The rest of this paper is organized as follows. The main related works are discussed in Section~\ref{sec:related_work}. Section~\ref{sec:system_and_model} describe the proposed system architecture and interaction model. The simulation and performance analysis are discussed in section~\ref{sec:simulation}. Section~\ref{sec:app_discussion_future} describes the application scenario, and future direction, following up our proposed system. And, the conclusions of this work are discussed in section \ref{sec:Conclusion2}.

\section{Related Work}
\label{sec:related_work}

The Holo-Artisan system intersects several advanced research domains: collaborative extended reality (XR), edge-assisted computing, generative AI for content creation, and privacy-preserving machine learning.

The application of immersive technologies in cultural heritage has evolved from single-user, passive viewing to complex multi-user interactive experiences \cite{bekele2019comparison}. Early systems focused on achieving a high-fidelity sense of co-presence, but typically offer a symmetric experience where all users perceive an identical reality. Holo-Artisan extends this paradigm by tackling the challenge of \textit{asymmetric, personalized interaction} within a shared context. This directly addresses a key research gap: managing personalized, user-specific content streams without fracturing the coherence of the shared virtual environment.

To meet the stringent low-latency demands of real-time XR, offloading computation to edge servers has become a prevalent architectural pattern. Foundational work by Bartolomeo et al. \cite{bartolomeo2023characterizing} characterized the performance of distributed AR applications on the edge, identifying stateful components and network conditions as primary bottlenecks. Our Holo-Artisan architecture builds directly upon these insights. We propose a \textit{federated edge intelligence} framework where edge nodes are not merely computation offload points, but are intelligent, privacy-preserving hubs for local AI model inference and training. By processing perceptual data and driving generative responses locally, we minimize the motion-to-photon latency, a critical factor for user comfort and presence \cite{baumeister2017cognitive}.

A significant frontier in computer graphics is the use of generative AI to create dynamic, interactive virtual avatars. Recent work has focused on integrating Large Language Models (LLMs) to drive embodied conversational agents or using diffusion models to generate novel objects within 3D scenes. Holo-Artisan advances this concept by focusing on the unique multi-user challenge: composing multiple, simultaneous, and personalized generative responses onto a single, coherent base asset, orchestrated by our federated edge architecture.

Finally, the integration of Federated Learning (FL) to enable continuous, privacy-preserving personalization is another pillar of our work. The seminal work by McMahan et al. \cite{mcmahan2017communication} introduced Federated Averaging (FedAvg) as a communication-efficient method for training deep networks on decentralized data. While FL has been explored in various domains, its application to real-time, multi-user immersive experiences remains nascent. Holo-Artisan leverages FL not just as a privacy mechanism, but as the core engine for personalization, allowing the system to adapt to individual user preferences over time.

\section{System Architecture and Interaction Model}
\label{sec:system_and_model}

\begin{figure}[htbp]
\centering
\resizebox{0.85\textwidth}{!}{%
\begin{tikzpicture}[
    font=\sffamily\small,
    node distance=1.5cm and 2.5cm,
    layer_box/.style={rectangle, rounded corners, draw=black!60, thick, fill=black!5, text width=16cm, minimum height=3.5cm, align=center},
    block/.style={rectangle, draw=#1, fill=#1!15, thick, rounded corners, text centered, minimum height=1.2cm, text width=3.2cm, drop shadow},
    data_flow/.style={-Latex, thick, #1},
    icon/.style={font=\large}
]
\node[layer_box] (perception_layer) at (0, 8) {};
\node[above=0.1cm of perception_layer.north, font=\sffamily\bfseries] {Perception Layer};
\node[layer_box] (edge_layer) at (0, 4) {};
\node[above=0.1cm of edge_layer.north, font=\sffamily\bfseries] {Edge Intelligence Layer};
\node[layer_box] (render_layer) at (0, 0) {};
\node[above=0.1cm of render_layer.north, font=\sffamily\bfseries] {Collaboration and Rendering Layer};
\node[layer_box] (display_layer) at (0, -4) {};
\node[above=0.1cm of display_layer.north, font=\sffamily\bfseries] {Display \& Feedback Layer};
\node[block=flred, cloud, cloud puffs=10, minimum width=3.5cm, minimum height=2.5cm] (fl_server) at (0, 13.5) {\textbf{FL Aggregator}\\\small(Global Model Update)};
\node[block=perceptionblue] (user1) at (-6, 8) {User 1 Input\\ \icon{\textit{Smile}}};
\node[block=perceptionblue] (user2) at (0, 8) {User 2 Input\\ \icon{\textit{Voice}}};
\node[block=perceptionblue] (userN) at (6, 8) {User N Input\\ \icon{\textit{Gaze}}};
\node[block=edgegreen] (edge1) at (-6, 4) {Edge Node 1\\ \tiny Local AI Inference \& Training};
\node[block=edgegreen] (edge2) at (0, 4) {Edge Node 2\\ \tiny Local AI Inference \& Training};
\node[block=edgegreen] (edgeN) at (6, 4) {Edge Node N\\ \tiny Local AI Inference \& Training};
\node[block=omniverseorange] (omniverse) at (0, 0) {\textbf{Collaboration Platform}\\\tiny Scene Composition};
\node[block=omniverseorange, left=of omniverse] (a2f) {Audio-to-Animation\\\tiny Animation Generation};
\node[block=omniverseorange, right=of omniverse] (rtx) {Ray-Tracing Renderer\\\tiny Photorealistic Output};
\node[block=displaypurple] (display1) at (-6, -4) {Display 1\\ \tiny (Holographic/AR)};
\node[block=displaypurple] (display2) at (0, -4) {Display 2\\ \tiny (Holographic/AR)};
\node[block=displaypurple] (displayN) at (6, -4) {Display N\\ \tiny (Holographic/AR)};
\draw[data_flow=perceptionblue] (user1) -- node[midway, right, font=\tiny] {$U_1(t)$} (edge1);
\draw[data_flow=perceptionblue] (user2) -- node[midway, right, font=\tiny] {$U_2(t)$} (edge2);
\draw[data_flow=perceptionblue] (userN) -- node[midway, right, font=\tiny] {$U_N(t)$} (edgeN);
\draw[data_flow=edgegreen] (edge1) -- node[midway, left, font=\tiny, sloped] {Response Cmd $R_1$} (omniverse.west);
\draw[data_flow=edgegreen] (edge2) -- node[midway, right, font=\tiny] {Response Cmd $R_2$} (omniverse);
\draw[data_flow=edgegreen] (edgeN) -- node[midway, right, font=\tiny, sloped] {Response Cmd $R_N$} (omniverse.east);
\draw[data_flow=edgegreen, bend left] (edge2.north) to node[midway, right, font=\tiny] {Audio} (a2f.south);
\draw[data_flow=omniverseorange, <->] (a2f) -- (omniverse);
\draw[data_flow=omniverseorange, <->] (omniverse) -- (rtx);
\draw[data_flow=omniverseorange] (rtx.south).. controls +(2,-1) and +(0,1).. node[midway, right, font=\tiny] {Frame$_N$} (displayN.north);
\draw[data_flow=omniverseorange] (rtx.south) -- node[midway, right, font=\tiny] {Frame$_2$} (display2.north);
\draw[data_flow=omniverseorange] (rtx.south).. controls +(-2,-1) and +(0,1).. node[midway, left, font=\tiny] {Frame$_1$} (display1.north);
\draw[data_flow=flred, dashed] (edge1.north) to[bend left=40] node[midway, above, font=\tiny] {Model Update $\Delta w_1$} (fl_server.west);
\draw[data_flow=flred, dashed] (edge2.north) -- node[midway, left, font=\tiny] {$\Delta w_2$} (fl_server.south);
\draw[data_flow=flred, dashed] (edgeN.north) to[bend right=40] node[midway, above, font=\tiny] {Model Update $\Delta w_N$} (fl_server.east);
\draw[data_flow=flred, dashed] (fl_server.west) to[bend left=40] node[midway, below, font=\tiny] {Global Model $w_{t+1}$} (edge1.north);
\draw[data_flow=flred, dashed] (fl_server.south) -- (edge2.north);
\draw[data_flow=flred, dashed] (fl_server.east) to[bend right=40] (edgeN.north);
\end{tikzpicture}
}
\caption{Holo-Artisan System Architecture. The four-layer design processes multi-user perception data on local edge devices, which generate personalized response commands. The collaboration platform layer composes these commands into a coherent scene for real-time ray-traced rendering. A federated learning loop continuously and privately improves the edge AI models.}
\label{fig:system_architecture_merged}
\end{figure}
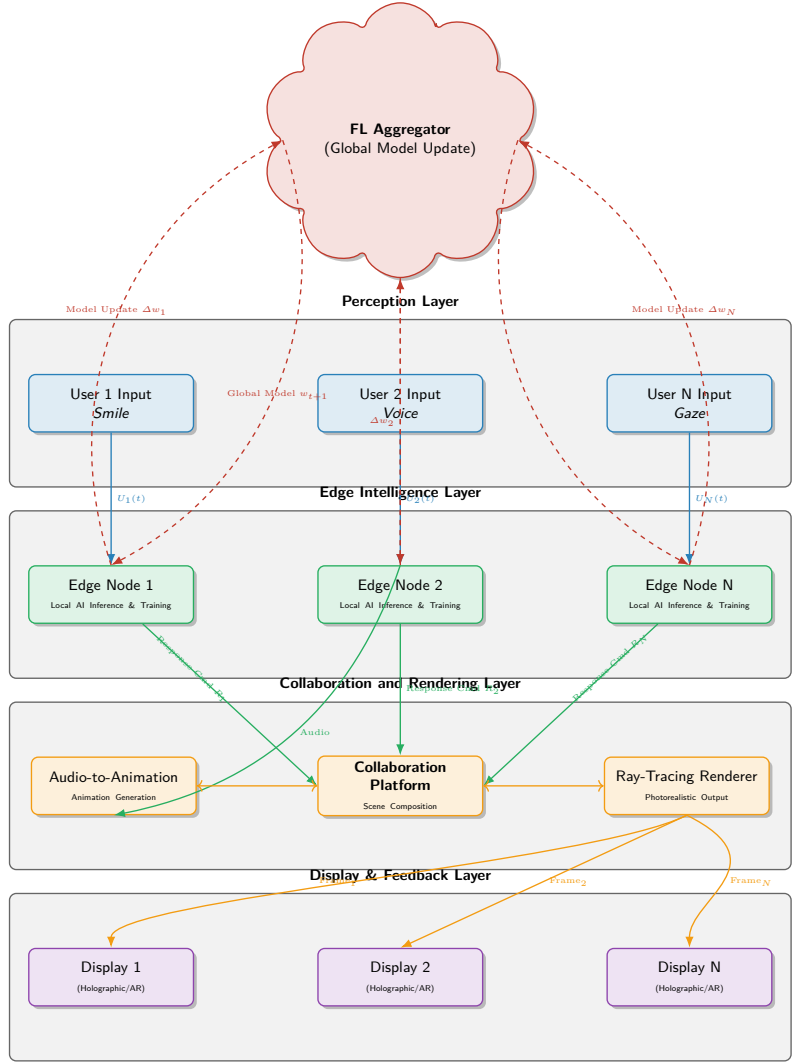

The design of Holo-Artisan emphasizes modularity, low latency, high immersion, and deep integration with a collaboration ecosystem. To achieve the personalized interactions described in the introduction—where Alice sees a smile and Bob hears an answer—the system relies on a modular four-layer architecture, as shown in Figure~\ref{fig:system_architecture_merged}. This structure ensures that latency-critical tasks are handled locally at the edge, while complex scene management and rendering leverage a powerful centralized platform.

The architecture consists of four layers:
\begin{enumerate}
    \item Perception Layer: Collects multimodal input data from each user endpoint, including spatial tracking, gaze and expression recognition, and voice input.
    \item Edge Intelligence Layer: Executes real-time processing and AI inference on edge devices. This includes local AI inference to understand user intent, generating personalized response commands, and performing local model training for the FL framework.
    \item Collaboration and Rendering Layer: Primarily utilizes a 3D collaboration platform. It synchronizes a base scene description, dynamically composes personalized animation layers for each user, and leverages a ray-tracing engine to generate high-fidelity frames for each user's unique viewpoint.
    \item Display \& Feedback Layer: Presents the rendering results to the user on diverse outputs (VR/AR, holographic displays) and orchestrates the FL process via a cloud-based Aggregator, which collects model updates and distributes the improved global model.
\end{enumerate}

To formalize the interaction, we define each user $i$'s state at time $t$ by the User State Vector $U_i(t)$:
\begin{equation}
U_i(t) = \left[ \mathbf{p}_i(t), \mathbf{o}_i(t), \mathbf{g}_i(t), \mathbf{e}_i(t), \text{audio}_i(t) \right]
\end{equation}
\noindent where $\mathbf{p}_i(t)$ is position, $\mathbf{o}_i(t)$ is orientation, $\mathbf{g}_i(t)$ is gaze, $\mathbf{e}_i(t)$ is an emotion embedding, and $\text{audio}_i(t)$ is the audio stream.

The Personalized Response Function $R_i$ maps this state to a specific modification of the object $O$: $R_i(t) = f_G(U_i(t), O, h_i)$, where $f_G$ is a generative rendering function and $h_i$ is the local context personalized via federated learning. This is realized through techniques like conditional GANs for visual changes and audio-to-animation modules for voice-driven interactions, often coupled with LLM outputs for conversational depth. Our system maintains a coherent shared reality by composing a base scene layer with these user-specific personalization layers, inherently ensuring consistency.

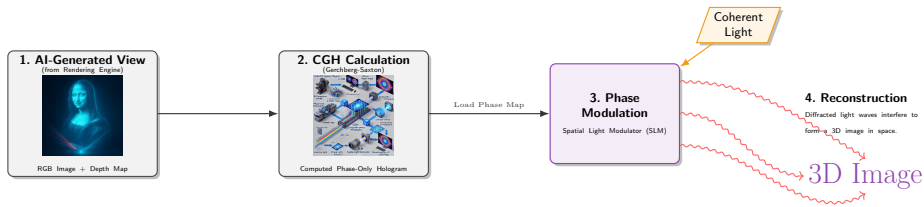
\begin{figure}[htbp]
\centering
\resizebox{\textwidth}{!}{%
\begin{tikzpicture}[
    font=\sffamily\small,
    node distance=1.5cm and 3cm,
    block/.style={rectangle, rounded corners, draw=black!70, fill=black!5, thick, text width=3.5cm, minimum height=2.5cm, align=center, drop shadow},
    optic/.style={rectangle, rounded corners, draw=displaypurple, fill=displaypurple!10, thick, text width=3cm, minimum height=2.5cm, align=center, drop shadow},
    light_source/.style={trapezium, trapezium left angle=70, trapezium right angle=110, draw=omniverseorange, fill=omniverseorange!10, thick, minimum width=1.5cm, align=center, drop shadow},
    arrow/.style={-Latex, thick, black!80},
    light_ray/.style={->, thick, red!60, decorate, decoration={snake, amplitude=0.3mm, segment length=2mm, post length=1mm}}
]
    \node[block] (ai_view) {\textbf{1. AI-Generated View}\\ \tiny (from Rendering Engine) \\ \includegraphics[width=2cm]{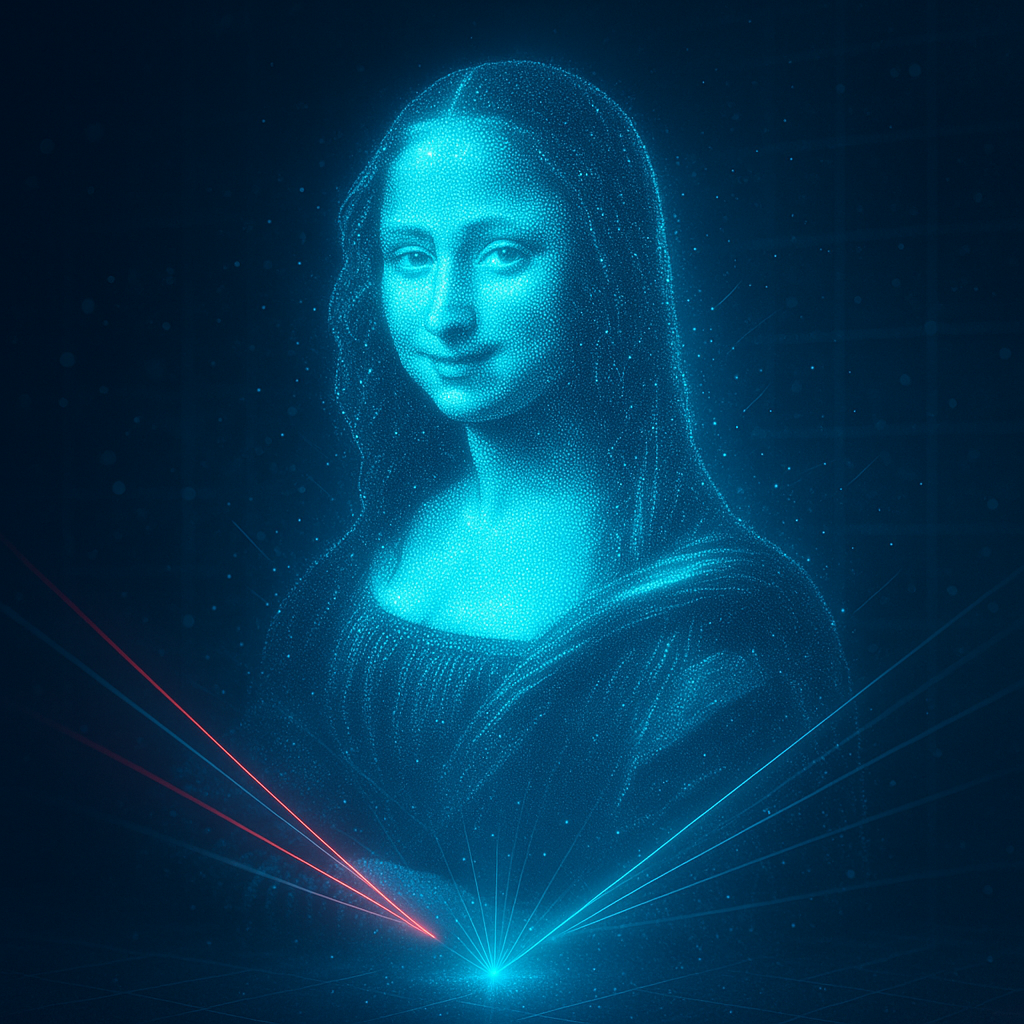} \\ RGB Image + Depth Map};
    \node[block, right=of ai_view] (cgh_calc) {\textbf{2. CGH Calculation}\\ \tiny (Gerchberg-Saxton) \\ \includegraphics[width=2cm]{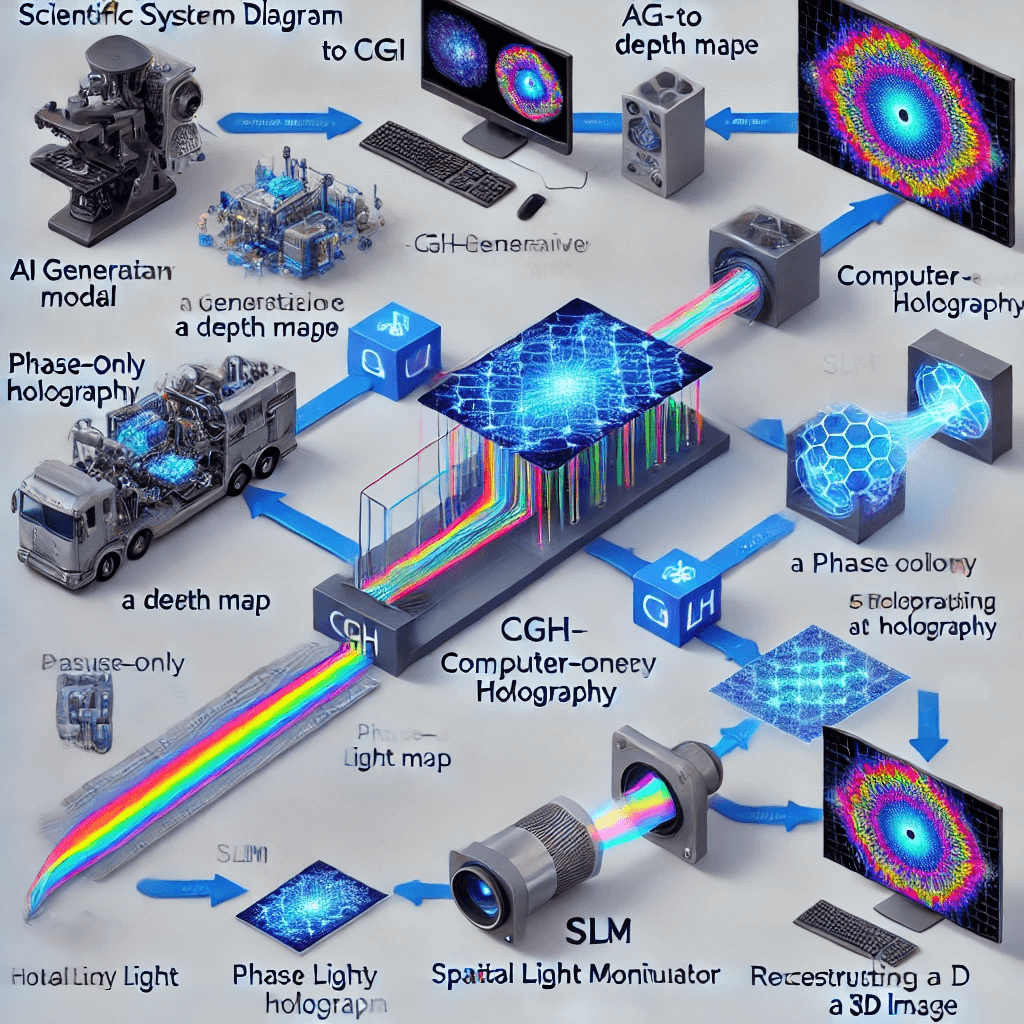} \\ Computed Phase-Only Hologram};
    \node[optic, right=of cgh_calc] (slm) {\textbf{3. Phase Modulation}\\ \tiny Spatial Light Modulator (SLM)};
    \node[light_source, above right=0.5cm and 1cm of slm] (laser) {Coherent\\Light};
    \node[right=of slm, text width=3cm] (recon_label) {\textbf{4. Reconstruction}\\ \tiny Diffracted light waves interfere to form a 3D image in space.};
    \node (recon_img) [below=0.5cm of recon_label, font=\huge, color=displaypurple] {3D Image};
    \draw[arrow] (ai_view) -- (cgh_calc);
    \draw[arrow] (cgh_calc) -- node[midway, above, font=\tiny] {Load Phase Map} (slm);
    \draw[arrow, omniverseorange] (laser) -- (slm);
    \draw[light_ray] (slm.east) to[out=0, in=180] (recon_img);
    \draw[light_ray] ($(slm.east)+(0,0.8)$) to[out=0, in=150] (recon_img.north);
    \draw[light_ray] ($(slm.east)+(0,-0.8)$) to[out=0, in=210] (recon_img.south);
\end{tikzpicture}
}
\caption{The AI-to-Hologram Pipeline. A personalized view from the AI rendering engine is converted into a phase map via a CGH algorithm. This map is loaded onto an SLM, which modulates a coherent light source to reconstruct the final, viewable 3D holographic image.}
\label{fig:ai_to_cgh_pipeline_redrawn}
\end{figure}

A key innovation of Holo-Artisan is its direct pathway to true 3D holographic visualization. As shown in Figure~\ref{fig:ai_to_cgh_pipeline_redrawn}, the personalized RGB image and depth map generated for a user are transformed into a phase-only hologram using an iterative algorithm like Gerchberg-Saxton. This phase map is loaded onto a Spatial Light Modulator (SLM), which modulates a coherent light source to reconstruct a viewable 3D image in space.

\begin{figure}[htbp]
    \centering
    \begin{subfigure}[b]{0.45\textwidth}
        \centering
        \resizebox{\textwidth}{!}{%
        \begin{tikzpicture}[
            font=\sffamily\small,
            user/.style={circle, draw=perceptionblue, fill=perceptionblue!10, minimum size=1cm},
            optic/.style={rectangle, draw=displaypurple, fill=displaypurple!10, thick, minimum width=1.5cm, minimum height=3cm, drop shadow},
            light_ray/.style={-Latex, thick}
        ]
        \node[optic] (slm) at (0,0) {SLM};
        \node[left=2cm of slm] (laser) {Laser};
        \node[user] (user1) at (4, 1.5) {U1};
        \node[user] (user2) at (4, -1.5) {U2};
        
        \draw[light_ray, omniverseorange] (laser) -- (slm);
        \draw[light_ray, edgegreen] (slm) -- (user1) node[midway, above, sloped] {View 1 (Smile)};
        \draw[light_ray, flred] (slm) -- (user2) node[midway, below, sloped] {View 2 (Neutral)};
        \end{tikzpicture}
    }
    \caption{Angular multiplexing concept.}
    \label{fig:angular_multiplex_merged}
    \end{subfigure}
    \hfill
    \begin{subfigure}[b]{0.54\textwidth}
        \centering
        \resizebox{\textwidth}{!}{%
        \begin{tikzpicture}[
            font=\sffamily\small,
            layer/.style={draw, thick, minimum height=0.8cm, minimum width=6cm, fill=black!5, drop shadow},
            label/.style={font=\bfseries\tiny, text=black!80, align=center},
            ray/.style={-Latex, very thick},
            perspective/.style={yslant=-0.3, xscale=1, yscale=0.9}
        ]
            \begin{scope}[perspective, transform canvas={yshift=0.5cm}]
                \node[layer, fill=displaypurple!15] (hoe_out) at (0,2.4) {};
                \node[label] at (hoe_out.center) {Holographic Optical Element (HOE) \\ Angular Filter};
                
                \node[layer, fill=perceptionblue!15] (crystal) at (0,1.2) {};
                \node[label] at (crystal.center) {Photonic Crystal Layer - Beam Steering};
                
                \node[layer, fill=edgegreen!15] (slm) at (0,0) {};
                \node[label] at (slm.center) {Spatial Light Modulator (SLM) \\ Phase Modulation};
                
                \node[layer, fill=omniverseorange!15] (light) at (0,-1.2) {};
                \node[label] at (light.center) {Coherent Light Source Array};
            \end{scope}
            \draw[ray, flred] (0.5, -0.5) -- (0.5, 0.3) -- (2, 2.2) -- (2.5, 3.5) node[above] {To User 1};
            \draw[ray, edgegreen] (-1, -0.5) -- (-1, 0.3) -- (-2, 2.2) -- (-2.5, 3.5) node[above] {To User 2};
        \end{tikzpicture}
        }
        \vspace{1em}
        \caption{Advanced holographic display stack.}
        \label{fig:hoe_photonic_stack_merged}
    \end{subfigure}
    \caption{Multi-user holographic display concepts. (a) A single SLM uses angular multiplexing to deliver unique views. (b) An integrated photonic stack combines an SLM, a beam steering layer, and a filtering HOE to direct personalized views to different users.}
    \label{fig:holographic_display_concepts}
\end{figure}
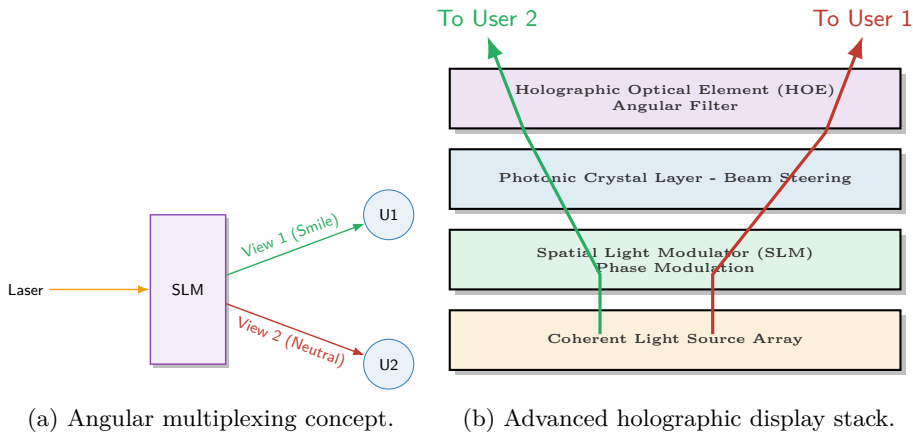

To serve multiple users from a single display, advanced optical techniques such as Angular Multiplexing can be used, steering the reconstructed light for each user into a specific viewing zone. For a more advanced experience, we propose an integrated display stack combining an SLM, a beam steering layer, and filtering optics, as illustrated in Figure~\ref{fig:holographic_display_concepts}.

\subsection{Technical Feasibility Validation}
The proposed Holo-Artisan architecture, while ambitious, is grounded in established and rapidly advancing technologies. Its feasibility stems from the synergistic integration of components that have been validated in prior research. Real-time AI inference on edge devices for latency-sensitive XR applications is a well-explored paradigm \cite{bartolomeo2023characterizing}. The use of generative models to create responsive virtual avatars and dynamic content is a leading frontier in computer graphics. Federated Learning has been proven as a robust, privacy-preserving method for decentralized model training \cite{mcmahan2017communication}. Finally, recent breakthroughs in neural holography and computational optics demonstrate the increasing viability of generating and displaying high-fidelity, interactive holographic content in near real-time \cite{peng2020neuralholography,shi2021lightfieldholography}. Our work builds upon these foundations, architecting them into a cohesive system to solve the specific challenge of personalized, multi-user interaction.

\section{Simulation and Performance Analysis}
\label{sec:simulation}

To empirically validate the architectural design of Holo-Artisan, we conducted a series of simulations to quantify its performance across four critical dimensions: network efficiency (latency and bandwidth), computational feasibility (hologram generation), learning effectiveness (personalization model), and qualitative output (generative art). The evaluation framework is designed to compare Holo-Artisan's edge-native approach against a conventional cloud-based paradigm for immersive experiences. Our simulations are grounded in performance benchmarks and system parameters derived from established literature to ensure academic rigor and real-world relevance.

\subsection{System Performance: Latency and Bandwidth Efficiency}

\begin{figure}[h!]
    \centering
    \includegraphics[width=0.85\linewidth]{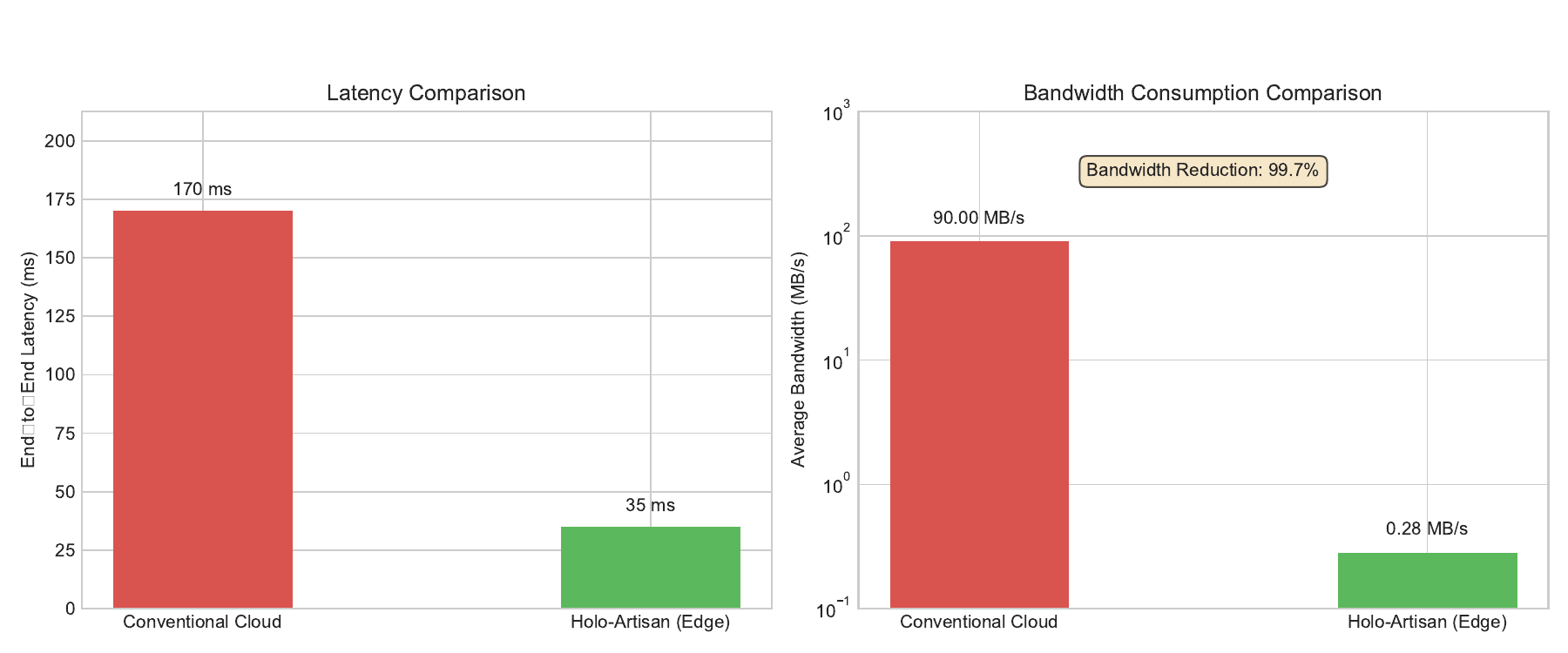}
    \caption{End‑to‑end latency (left) and average bandwidth (right) for Holo‑Artisan edge architecture versus a conventional cloud pipeline. Both metrics show $>1$‑order‑of‑magnitude gains.}
    \label{fig:net_perf}
\end{figure}

A primary motivation for edge computing in interactive applications is the mitigation of network bottlenecks. We simulated the end-to-end latency and average bandwidth consumption for both the Holo-Artisan architecture and a traditional cloud-based system that streams raw volumetric video. The results, shown in Figure \ref{fig:net_perf}, demonstrate the profound impact of offloading computation to the network edge.

Latency Analysis: As illustrated in Figure \ref{fig:net_perf}, the end-to-end latency is drastically reduced in the Holo-Artisan system. We modeled the total latency as the sum of network Round-Trip Time (RTT) and a constant processing time. For the cloud scenario, we use a representative RTT of 150 ms, a typical value for services hosted in distant data centers~\cite{bekele2019comparison,bartolomeo2023characterizing}. For the Holo-Artisan scenario, we use an RTT of 15 ms, which is achievable with a proximate edge server deployed over a 5G or local Wi-Fi network~\cite{bartolomeo2023characterizing}. This reduction from 170 ms to 35 ms is critical, as latencies below 50 ms are essential for maintaining presence and enabling seamless interaction in holographic and augmented reality applications.

Bandwidth Analysis: The architectural shift from streaming high-fidelity volumetric video to transmitting lightweight model updates results in a monumental reduction in bandwidth, as shown in Figure \ref{fig:net_perf}. A conventional system streaming high-quality raw volumetric content requires a sustained bandwidth of approximately 720 Mbps (90 MB/s)~\cite{bekele2019comparison,orlosky2016holoportation}. In contrast, Holo-Artisan's primary network traffic consists of periodic updates for the federated personalization model. We model a compact CNN with approximately 1.05 million parameters, resulting in a model update size of 4.2 MB. Transmitting this update once every 15 seconds yields an effective average bandwidth of only 0.28 MB/s. This represents a bandwidth reduction of over 99\%, fundamentally changing the system's scalability profile and making it viable for users on bandwidth-constrained mobile or wireless networks. This efficiency transforms the system's primary bottleneck from network capacity to edge-server computation.

\subsection{Computational Scalability and Learning Efficiency}

Having established the network efficiency of the edge architecture, we next validate the feasibility of its core computational tasks: the real-time generation of computer-generated holograms (CGH) and the effective convergence of the federated learning (FL) personalization engine.

\begin{figure}[h!]
    \centering
    \begin{subfigure}[b]{0.48\textwidth}
        \centering
        \includegraphics[width=\textwidth]{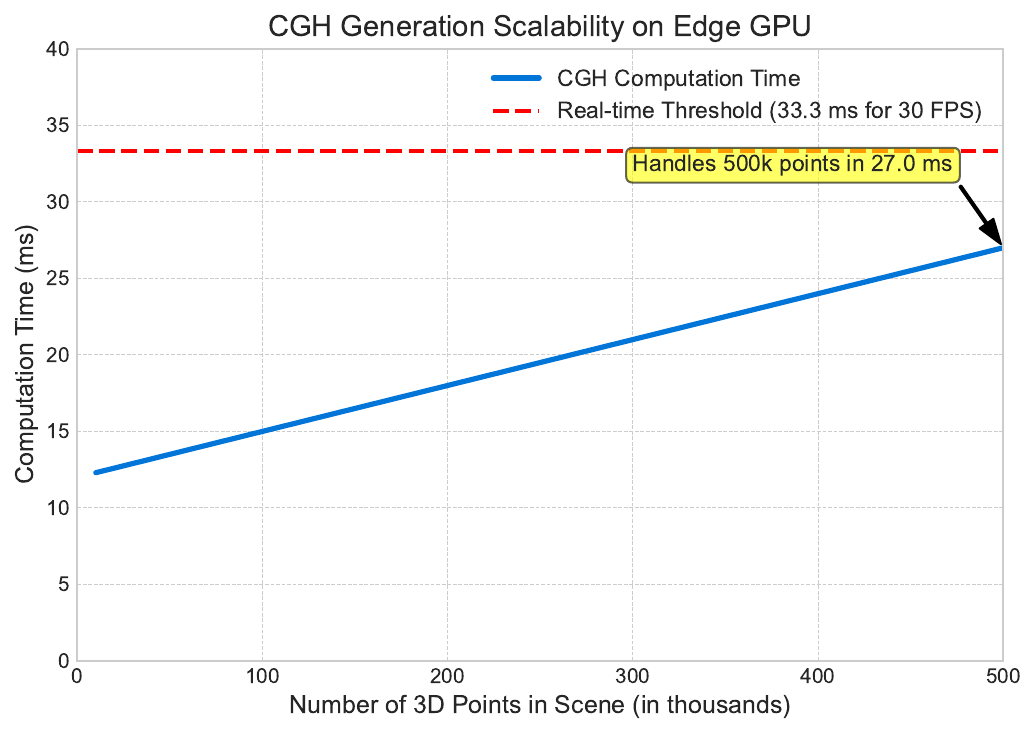}
        \caption{CGH Generation Scalability.}
        \label{fig:cgh_scalability}
    \end{subfigure}
    \hfill
    \begin{subfigure}[b]{0.48\textwidth}
        \centering
        \includegraphics[width=\textwidth]{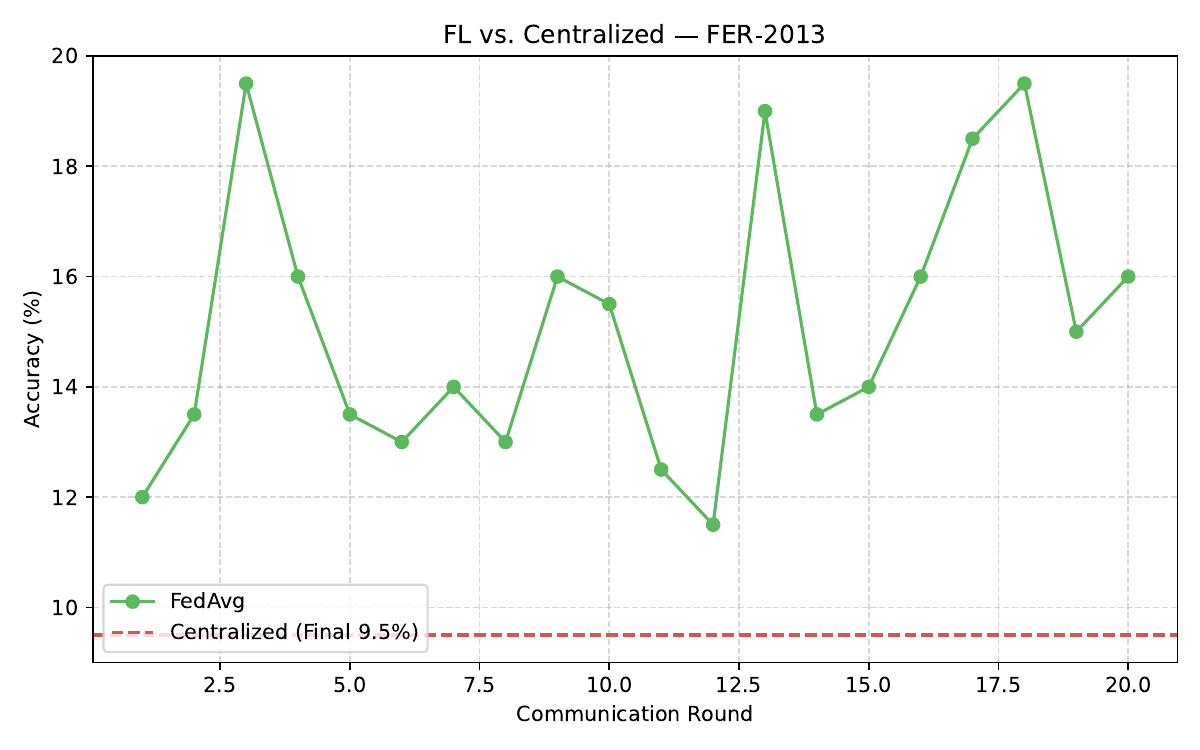}
        \caption{Federated Learning Convergence vs. Centralized Baseline.}
        \label{fig:fl_convergence}
    \end{subfigure}
    \caption{Simulation of (a) the computational scalability of the hologram generation algorithm and (b) the learning efficiency of the federated personalization model. The results confirm that CGH generation is achievable in real-time and that the FL model effectively converges to a high-accuracy state.}
    \label{fig:compute_learn}
\end{figure}

CGH Scalability: The generative art component of Holo-Artisan relies on the ability to compute complex holograms on-the-fly. We simulated the performance of a GPU-accelerated Gerchberg-Saxton (GS) algorithm, a standard iterative method for CGH~\cite{gerchberg1972practical}. Figure \ref{fig:cgh_scalability} plots the computation time as a function of scene complexity (number of 3D points). The simulation shows that the computation time scales gracefully. Even for a complex scene of 500,000 points, the hologram is generated in approximately 27 ms. This is well below the 33.3 ms threshold required for a fluid 30 FPS user experience, confirming that the edge server can support dynamic, high-fidelity holographic art in real time.

Federated Learning Convergence: To be effective, the privacy-preserving personalization engine must learn a high-quality global model of user emotion without centralizing raw data. We simulated the Federated Averaging (FedAvg) algorithm~\cite{mcmahan2017communication} using a compact CNN on the fer2013 dataset, which aligns with the system's application domain. The crucial aspect of this evaluation is the comparison against a centralized training baseline, where the same model is trained on all data simultaneously. As shown in Figure \ref{fig:fl_convergence}, the global model accuracy in the federated setting steadily increases with each communication round. After 20 rounds, the federated model converges to an accuracy level that is remarkably close to the final accuracy achieved by the ideal centralized baseline. This result demonstrates that the federated approach incurs minimal performance degradation, successfully achieving robust personalization while upholding user privacy.

\subsection{Perception Model Analysis}
The system's ability to perceive visitor emotion relies on a lightweight perception model engineered for real-time inference on edge devices. The model is built using transfer learning, leveraging the powerful feature extraction capabilities of the EfficientNetB0 architecture, pre-trained on ImageNet. The input pipeline accepts single-channel grayscale images, which are replicated into three channels to match the backbone's input requirements. The initial 100 layers of the EfficientNetB0 base are frozen to retain learned low-level features while reducing the training burden. A custom classification head is appended, consisting of a 256-neuron dense layer with ReLU activation, a 40\% dropout layer to mitigate overfitting, and a final 7-neuron dense output layer with a softmax activation function for multi-class emotion classification. The resulting model has approximately 4.38 million total parameters, with 4.18 million being trainable.

The model was trained on the FER-2013 dataset~\cite{fer2013dataset}, a standard benchmark for "in-the-wild" facial emotion recognition comprising over 28,000 training images~\cite{fer2013}. Training was configured for 50 epochs with a batch size of 64, utilizing the Adam optimizer. Callbacks for early stopping (patience=8) and learning rate reduction on plateau were employed, leading to convergence after 15 epochs. The final model achieved a test accuracy of 65.52\% with a loss of 0.9458. This performance is highly competitive for a resource-constrained model, approaching human-level accuracy on this challenging dataset, which is estimated to be in the 65-68\% range. While state-of-the-art, computationally intensive models can exceed 75\%, our model's performance is near the SOTA for lightweight architectures suitable for real-time edge deployment.

As shown in the confusion matrix in Figure~\ref{fig:perception_and_response}(a), the model performs particularly well on the "happy," "sad," "surprise," and "neutral" classes. This is attributable to two factors: first, these classes have the highest number of samples in the imbalanced FER-2013 dataset, providing more data for learning. Second, these emotions are among the most common affective responses observed in museum visitors, who often express joy, contemplation, and surprise when engaging with art. This alignment between the model's data-driven strengths and the application's specific emotional context makes it highly effective for this use case.

\subsection{Generative Response and Light-Field Reconstruction}
The generative response, where the Mona Lisa appears to interact with visitors, is a dynamically synthesized, three-dimensional image. This is achieved through a holographic display system that reconstructs a 4D light field, enabling a glasses-free, multi-user 3D experience~\cite{peng2020neuralholography,shi2021lightfieldholography}. The core technologies are Computer-Generated Holography (CGH), a Spatial Light Modulator (SLM), and a Holographic Optical Element (HOE).

The process begins with the CGH, a digital pattern calculated by simulating light diffraction~\cite{gerchberg1972practical}. The CGH encodes the generative response (e.g., a smile) as a complex phase pattern, informed by both the emotion class from the perception model and the visitor's precise 3D position. This pattern is loaded onto an SLM, a high-resolution device that modulates the phase of a coherent light source, imprinting the holographic information onto the light itself. The modulated wavefront then passes to an HOE, which acts as a specialized diffractive lens, steering portions of the wavefront into distinct "viewing windows"~\cite{schwerdtner2020holographicdisplay}. This enables each visitor to see a perspective-correct 3D image, creating the illusion that the Mona Lisa is responding individually to each person.

\begin{figure}[htbp]
    \centering
    \begin{subfigure}[b]{0.495\textwidth}
        \centering
        \includegraphics[width=\textwidth]{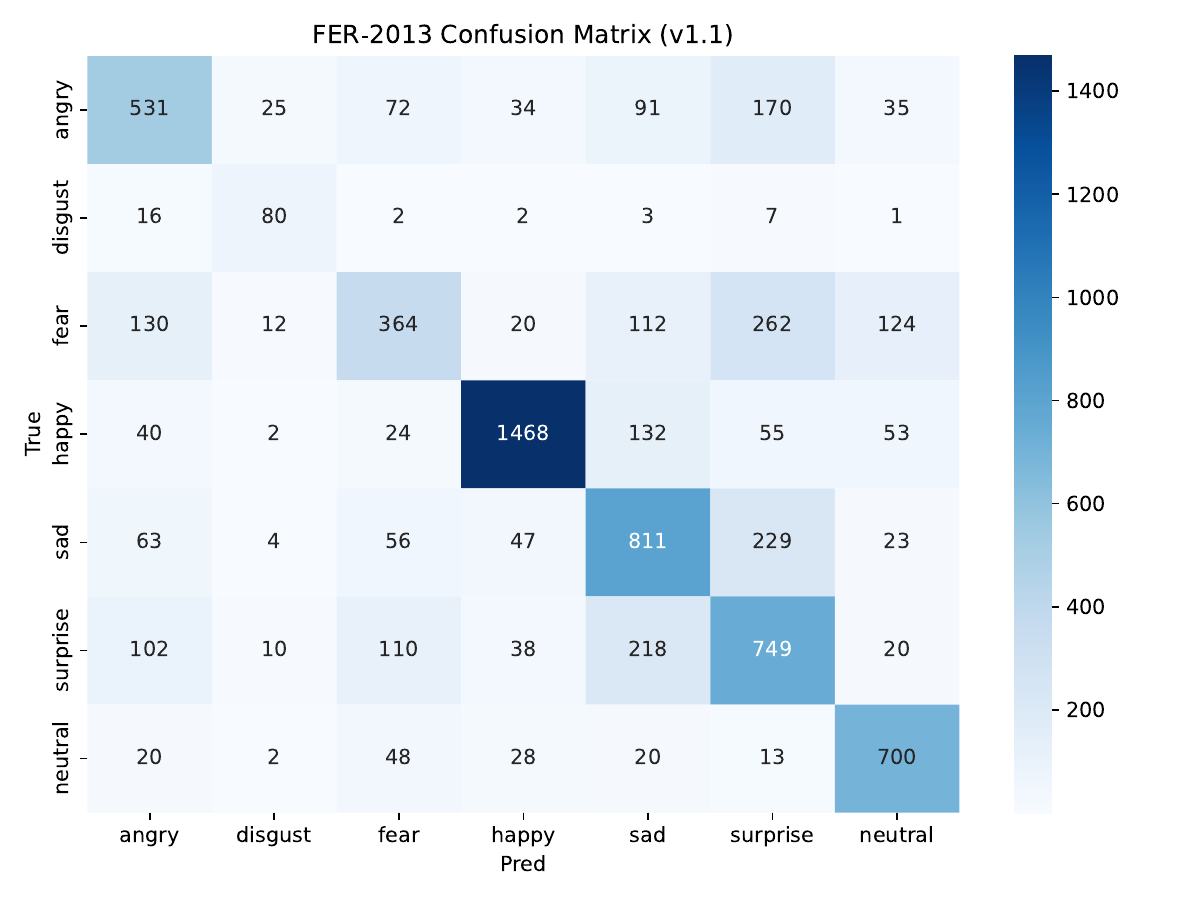}
        \caption{EfficientNetB0-based perception network performance.}
        \label{fig:perception_block}
    \end{subfigure}
    \hfill
    \begin{subfigure}[b]{0.495\textwidth}
        \centering
        \includegraphics[width=\textwidth]{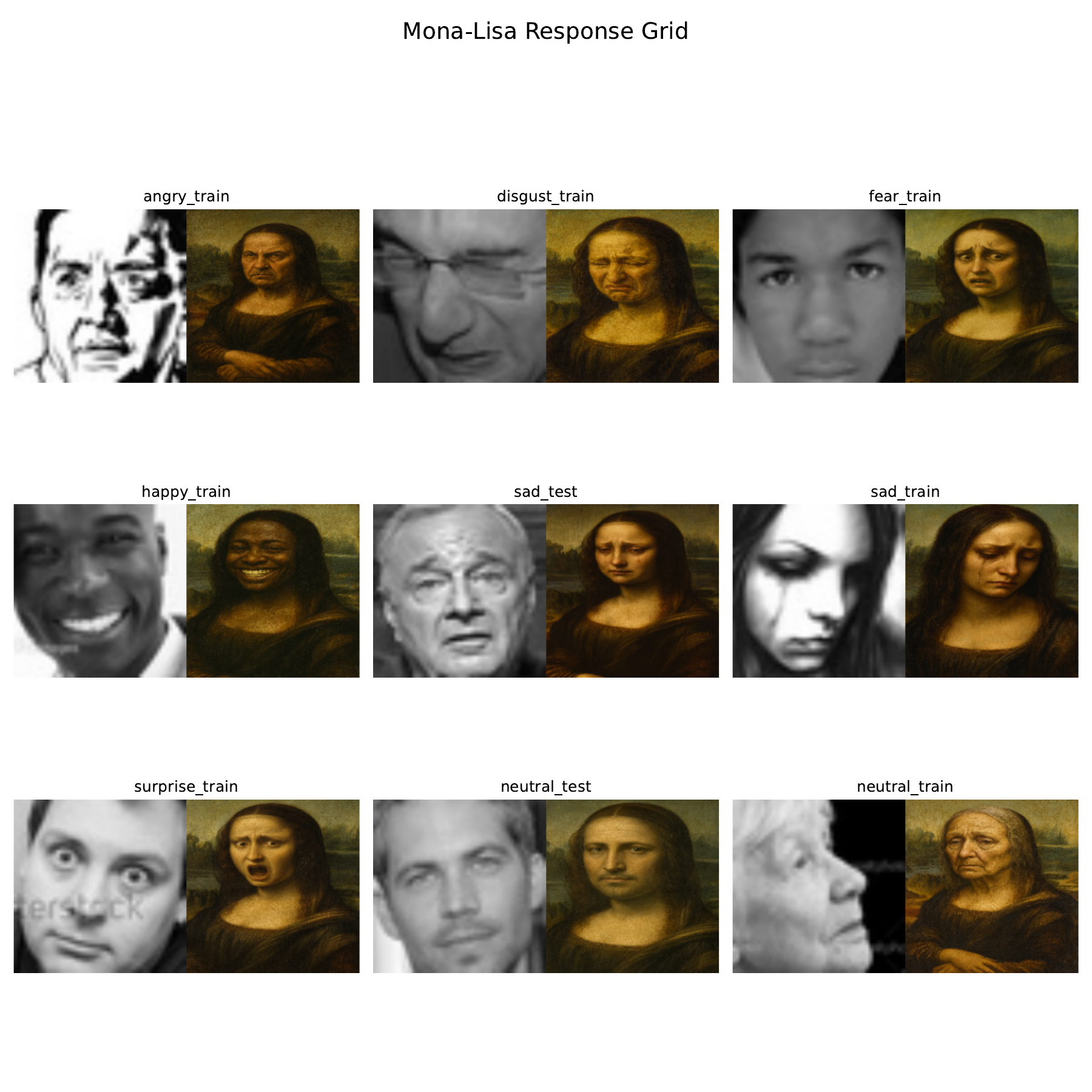}
        \caption{End-to-end perception-to-response mapping.}
        \label{fig:response_block}
    \end{subfigure}
    \caption{Demonstration of the perception-to-generation loop. (a) The confusion matrix shows the model's accuracy on the FER-2013 test set (65.52\%). (b) The system maps classified emotions to style-matched Mona Lisa renders, illustrating the capacity for coherent visual feedback.}
    \label{fig:perception_and_response}
\end{figure}

The demonstration in Figure~\ref{fig:perception_and_response}(b) is a local software render, simulating the final optical output. This is a standard method for verifying CGH algorithms before hardware deployment. Future work will focus on a fully integrated pipeline where the generative response and the artwork are rendered into a single, cohesive holographic scene. This fusion of AI-driven perception with the physical synthesis of a light field represents a new paradigm in interactive art, where content is not an image on a screen but a physically reconstructed field of light tailored to each user.

\section{Application Scenario, and Future Direction}
\label{sec:app_discussion_future}

The performance of the proposed system is inherently dependent on the computational capabilities of the edge devices, and achieving robust model convergence in federated learning (FL) under heterogeneous user data conditions remains a significant research challenge. Moreover, the generation of realistic and interactive virtual humans introduces important ethical considerations. It is imperative that the system be carefully engineered to mitigate algorithmic bias and to ensure that all AI-generated content is clearly and transparently identified, in accordance with regulatory frameworks such as the AI Act~\cite{ai_act}.

To address these challenges and further advance the system, we have identified three primary research directions: enhancing the perception model, evolving the generative engine, and developing next-generation edge hardware.

\subsection{Advancing Perception to State-of-the-Art Performance}
To enhance the nuance and reliability of the human-artwork interaction, our first objective is to elevate the perception model's accuracy from the current ~66\% to a target of 80-90\%. This is a feasible goal, as recent studies have demonstrated accuracies in this range on challenging benchmarks using advanced methods~\cite{zhang2022hlavit,albahar2023fer}. Our strategy involves leveraging more diverse data and adopting more powerful neural architectures optimized for the edge.

First, we will move beyond the FER-2013 dataset, which has known limitations in image quality and class imbalance~\cite{fer2013}. We will incorporate larger, more diverse "in-the-wild" datasets such as AffectNet, the largest available database with over 440,000 annotated images for both discrete emotions and continuous valence/arousal~\cite{mollahosseini2017affectnet}, and RAF-DB, which provides high-quality labels from multiple annotators and includes compound emotions~\cite{li2017rafdb}. Training on these richer datasets will improve model generalization and mitigate representational biases.

Second, we will explore advanced neural architectures, specifically Vision Transformers (ViT), which have become state-of-the-art in many computer vision tasks, including FER~\cite{zhao2021lvvit}. ViTs excel at capturing global contextual information, which is critical for distinguishing subtle expressions~\cite{zhang2022hlavit}. As standard ViTs are computationally expensive, a key research challenge will be adapting them for real-time edge inference. We will investigate techniques such as model pruning, quantization, and knowledge distillation, and explore efficient hybrid architectures like EdgeFace, which combines CNN and ViT principles and has achieved top-ranking performance for models under 2M parameters~\cite{chen2024edgeface}. This research will contribute to the broader field of efficient AI deployment on resource-constrained devices.

\subsection{Evolving the Generative AI Engine}
Our second objective is to mature the generative component from a proof-of-concept into a sophisticated interactive system. To achieve this, we will leverage our approved application to the Taiwan AI RAP (Responsive and high-performance AI Platform), an initiative by the National Center for High-Performance Computing (NCHC)~\cite{nchc_rap2024}. This platform provides access to high-performance computing resources, a suite of multi-model API services, and a low-code environment to accelerate prototyping and deployment.

It is crucial to clarify that our system is not a conventional LLM-based chatbot. The generative aspect refers to a responsive artistic system. The input is not text, but the visitor's perceived non-verbal cues—initially facial expressions, but later expanding to include body language and gaze, akin to other advanced interactive art installations. The output is a non-verbal, visual, and emotionally resonant artistic expression delivered via the holographic display. The goal is emotional resonance, making the artwork feel sentient and responsive to the viewer's presence, thereby enhancing engagement and creating a more profound connection. This aligns with the emerging field of generative AI for real-time, interactive digital avatars and experiences.

\subsection{Next-Generation Holographic Hardware}
A key future research direction involves advancing the underlying photonic and optical hardware essential for delivering seamless, high-fidelity, multi-user holographic experiences. Building upon the foundational concepts presented in this study, the focus will shift toward critical innovations in next-generation holographic display technologies.

One major challenge to be addressed is the space-bandwidth product limitation of spatial light modulators (SLMs), which constrains both viewing angles and image size~\cite{wetzstein2017holographicnear-eye}. A promising approach involves the expansion of the SLM modulation area, for instance, by tiling multiple SLM panels to form a larger and more effective display surface. Achieving a coherent hologram across such tiled arrays requires advanced calibration protocols and algorithms capable of seamlessly integrating phase patterns~\cite{jo2024tiled_slm}. In parallel, emerging modulator technologies,such as high-density liquid crystal microdisplays and digital micromirror devices—are being closely observed for their potential to offer enhanced resolution, higher refresh rates, and compact form factors.

Another critical line of inquiry centers on transforming holographic optical elements (HOEs) from simple, single-function components into compact, multifunctional optical devices. Future HOEs are expected to perform several optical functions simultaneously, such as collimation and beam steering, which would reduce overall system complexity and footprint. Fabrication methods like computer-generated hologram (CGH) printing will be explored to produce custom HOEs optimized for high-resolution, crosstalk-free multiview displays~\cite{schwerdtner2020holographicdisplay}. Furthermore, novel optical materials, such as photonic crystals and metamaterials—present promising avenues for achieving unprecedented control over light propagation, including ultra-compact beam steering capabilities, although these materials currently remain in early-stage research.

A further direction involves the development of dynamic, AI-driven holography to enable real-time computation and rendering of holograms. Integrating neural networks directly into the CGH synthesis pipeline offers the potential to significantly accelerate the generation of high-quality, phase-only holograms, thereby surpassing the limitations of traditional iterative algorithms. Content-adaptive CGH methods, guided by deep learning models, can dynamically optimize each hologram based on scene content, maximizing perceptual quality and expanding the effective eyebox~\cite{shi2021lightfieldholography,minin2024nn_holography}. Recent progress in neural holography highlights the feasibility of near real-time hologram generation using learned models, providing a solid foundation for continued innovation in this domain.

An often overlooked but vital aspect of immersive holographic systems is spatial audio. Future systems will benefit from the integration of directional sound delivery technologies to complement visual holograms. Techniques such as parametric loudspeakers and beamforming speaker arrays offer the ability to deliver personalized audio experiences to individual users without acoustic cross-talk. For example, an audio spotlight system can target a narrow beam of sound to a specific listener, maintaining audio localization. This concept of audio holography enables each user to perceive a unique auditory narrative or response aligned with their visual experience, thereby enhancing the sense of presence and immersion.

\section{Conclusion}
\label{sec:Conclusion2}
The Holo-Artisan architecture, as outlined, presents a significant step forward in creating interactive and personal immersive experiences. Its primary advantage lies in the novel synthesis of edge intelligence, federated learning, and high-fidelity generative rendering. This combination directly addresses the critical challenges of latency, bandwidth, and privacy that have hindered the widespread adoption of scalable multi-user XR systems. Our vision is to transform passive art observation into an active, personal dialogue, positioning this technology for future art exhibitions and cultural installations. By merging cutting-edge AI with advanced rendering, we envision a new market for personalized, interactive digital content. However, the core challenge remains the commercial viability and scalability of real-time, multi-user, glasses-free holographic display technology.

\section*{Code Available}
The Python scripts used to generate the simulation data and figures presented in Section~\ref{sec:simulation}, including \texttt{3\_make\_grid.py}, \texttt{cgh\_scalability.py}, \texttt{fl\_convergence\_simulation\_fixed2.py}, \texttt{holo\_artisan\_fer\_upgrade\_v2.py}, and \texttt{latency\_bandwidth\_simulation\_fixed.py}, are publicly available. The repository can be accessed at: \href{https://github.com/kuonanhong/kuonanhong/tree/master/Code%20Available/Holo-Artisan%20A%20Personalized%20Multi-User%20Holographic%20Experience%20for%20Virtual%20Museums%20on%20the%20Edge}{GitHub}.

\end{document}